\documentclass[twoside,slac_one]{revtex4}
\usepackage{graphicx}
\usepackage{fancyhdr}
\usepackage{amsmath} 
\usepackage{bm}
\usepackage{amsxtra}
\usepackage{amssymb}
\usepackage{amsthm}
\usepackage{latexsym}
\usepackage{lscape}

\begin{document}

\title{SciNO$\nu$A: A measurement of neutrino-nucleus scattering in a narrow-band beam}

%

\author{X. C. Tian, on behalf of the SciNO$\nu$A Study Group}
\affiliation{Department of Physics and Astronomy, University of South Carolina, Columbia, SC, USA}

\begin{abstract}
SciNO$\nu$A is a proposed experiment to deploy a fine-grained scintillator detector
in front of the NO$\nu$A near detector to collect neutrino-nucleus scattering events
in the NuMI, off-axis, narrow-band neutrino beam at Fermilab. This detector
can make unique contributions to the measurement of charged- and neutral-current
quasi-elastic scattering; and neutral-current $\pi^0$ and photon production. These
processes are important to understand for fundamental physics and as backgrounds
to measurements of electron neutrino appearance oscillations. 
\end{abstract}

\maketitle

\thispagestyle{fancy}


\section{Introduction}
\begin{figure}[h]
\centering
\includegraphics[width=80mm]{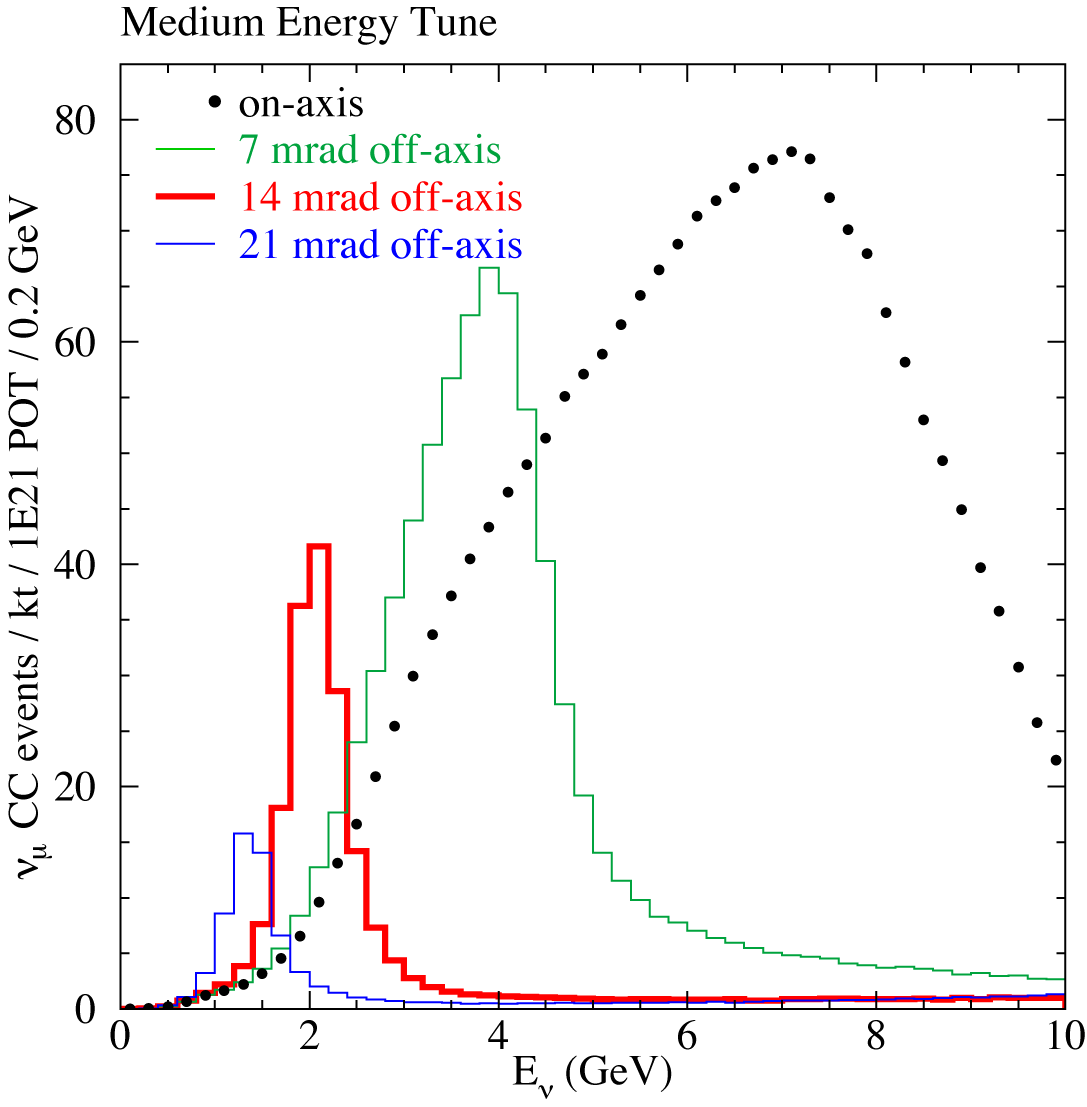}
\caption{The NuMI neutrino energy spectrum at different off-axis angles. The NO$\nu$A detectors are located at 14 mrad.} \label{beam}
\end{figure}

The next generation of long baseline neutrino experiments aims to measure the third mixing angle $\theta_{13}$, 
determine whether CP is violated in the lepton sector, and resolve the neutrino mass hierarchy~\cite{NOvA, T2K, LBNE}. The NuMI Off-axis 
electron-neutrino ($\nu_e$) Appearance (NO$\nu$A) experiment is the flagship experiment of the US domestic 
particle physics program which has the potential to address most of the fundamental questions in neutrino physics 
raised by the Particle Physics Project Prioritization Panel (P5). 
NO$\nu$A has two detectors, a 222 ton near detector located underground at Fermilab and a 14 kiloton far 
detector located in Ash River, Minnesota with a baseline of 810 km. The detectors are composed of extruded PVC 
cells loaded with Titanium dioxide to enhance reflectivity. There are 16,416 and 356,352 cells for the near and far 
detector, respectively. Each cell has an size of 3.93 cm transverse to the beam direction and 6.12 cm along 
the direction. The corresponding radiation length is 0.15 $X_0$, ideal for the identification of electron-type neutrino 
events. The ``Neutrinos at the Main Injector'' (NuMI) beam will provide a 14 mrad off-axis neutrino beam to reduce neutral 
current backgrounds and which peaks at 2 GeV, corresponding to the first oscillation maximum for this detector 
distance as shown in Fig.~\ref{beam}. The accelerator and NuMI upgrades will double the protons per year delivered 
to the detector which is $6\times 10^{20}$ protons per year. 

The interpretation of the oscillation results requires detailed knowledge of the neutrino spectrum, neutrino cross-sections,
nuclear effects, and final state topologies. SciNO$\nu$A is a proposed fine-grained detector ~\cite{SciNOvA} placed directly 
in front of the NO$\nu$A near detector which can be used to measure neutrino-nucleus scattering.  
Comparing with the wide band beams used by the MINER$\nu$A and SciBooNE experiments, the narrow band beam will provide 
better apriori knowledge of the incident neutrino energy, and  has lower background from high energy feed-down. In the 
energy region around 2 GeV, the SciNO$\nu$A detector will record about 1 M events per year as shown in Tab.~\ref{statistics}.
\begin{table}[h]
\begin{center}
\caption{The event rates per year for elastic, resonant, deep-inelastic (DIS) and coherent neutrino scattering recorded by 
SciNO$\nu$A detector calculated using the GENIE neutrino generator~\cite{GENIE}.}
\begin{tabular}{|l|c|c|c|}
\hline \textbf{}                                          & \textbf{Charged-Current} ($\times 10^3$) & \textbf{Neutral-Current} ($\times 10^3$) \\
\hline Elastic                                           & 220 & 86 \\
\hline Resonant                                     & 327 & 115 \\
\hline DIS                                                & 289 & 96 \\
\hline Coherent                                      & 8     & 5     \\
\hline Total                                              & 845 & 302 \\
\hline $\nu + A \rightarrow \pi^0 + X$ & 204 & 106 \\
\hline
\end{tabular}
\label{statistics}
\end{center}
\end{table}

\section{SciNO$\nu$A Detector}
\begin{figure}[h]
\centering
\includegraphics[width=0.80\textwidth]{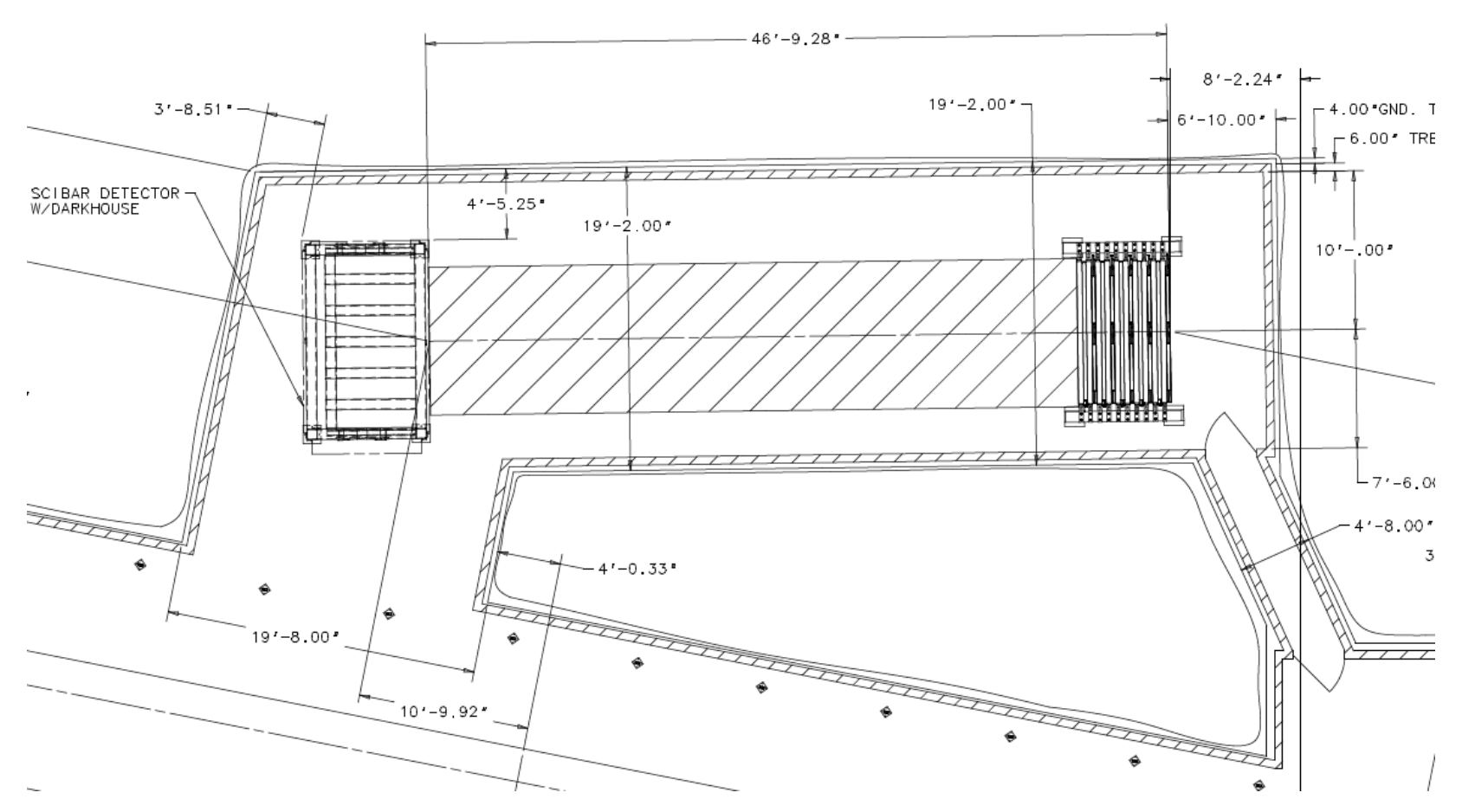}
\caption{The layout of the SciNO$\nu$A detector and the NO$\nu$A near detector.} \label{scibar}
\end{figure}

The baseline design of the SciNO$\nu$A detector is a copy of the SciBar detector used as the near detector for the K2K experiment~\cite{K2K-scibar} 
in Japan and SciBooNE experiment~\cite{SciBooNE} at Fermilab. The detector is composed of 14,848 extruded 1.3$\times$2.5$\times$290 cm$^3$ 
scintillator strips arranged in 64 layers. Each layer consists of an $X$ and $Y$  plane enabling a 3-D reconstruction, each containing 
116 strips. The scintillator strips are made of polystyrene doped with PPO and POPOP and are co-extruded with a TiO$_2$ reflection 
coating 0.25 mm thick to increase the reflectivity.  The total active volume of the scintillator is 2.9$\times$2.9$\times$ 1.7 m$^3$ with 
total mass of 15 tons. Each strip contains an embedded 1.5 mm diameter wavelength-shifting (WLS) fiber used to collect the light from 
the scintillator extrusions. A group of 64 fibers will be gathered into a 64-anode multianode photomultiplier tubes (MAPMT) for readout. 
The Hamamatsu H8804 MAPMTs will be used as the photo detector. It has a bialkali cathode yielding a quantum efficiency of $\sim12\%$ 
at the WLS fiber emission wavelength around 480 nm. At 800V, the MAPMTs will provide a gain of $6\times10^5$ with typical pixel-pixel 
uniformity of 1:2.5 and cross talk of 4\% for adjacent pixels. A 12-PMT prototype readout system has been built and is running at Indiana 
University which can be adopted to be used as the SciNO$\nu$A detector readout system for the 232 PMTs. Other technology options, 
such as the size of the scintillator strips, the photo detector technologies, and corresponding readout system are under study in order to 
optimize the performance of the SciNO$\nu$A detector. The steel detector frame used by the SciBooNE experiment will be reused as 
the detector support structure to save cost. The SciNO$\nu$A detector will be placed directly in front of the NO$\nu$A near detector 
as shown in Fig.~\ref{scibar}. There is adequate space in front of the NO$\nu$A near detector to accommodate the SciNO$\nu$A detector, so no 
additional cavern excavation will be needed. 

The total cost of the proposed SciNO$\nu$A detector is \$2.4 M (baseline design), dominated by the detector construction and assembly (\$0.8 M) 
and new readout electronics (\$1.42 M). The expected start time is the summer of 2012 and will be ready when the NO$\nu$A Far Detector is 
ready in early 2014. The SciNO$\nu$A proposal was presented to the Fermilab Physics Advisory Committee (PAC) in  November 2010 and 
the PAC recommended that NO$\nu$A consider SciNO$\nu$A.  The NO$\nu$A collaboration supports the SciNO$\nu$A physics case and is 
seriously evaluating it as a possibility. A study group consisting of NO$\nu$A and non-NO$\nu$A physicists recently formed to answer remaining 
technical questions. The funding is correctly being sougth.

\section{SciNO$\nu$A Physics}
The fine-grained SciNO$\nu$A detector using the narrow band beam will accumulate around 1 M events per year as shown in 
Tab. ~\ref{statistics} which will enable a detailed study of neutrino-nucleus scattering. This not only will provide a significant 
cross check of NO$\nu$A oscillation background processes, but also will add important information to better understand some 
current mysteries, including the anomalously large charged-current quasi-elastic cross section measured by MiniBooNE~\cite{MiniBooNE-CCQE} 
and the apparent absence of charged current coherent pion production observed by the K2K~\cite{K2K-CCCo} and SciBooNE~\cite{SciBooNE-CCCo} 
experiments at low neutrino energies. SciNO$\nu$A may also be able to shed further light on the excess of electron-like events observed 
in MiniBooNE~\cite{MiniBooNE-Nue,Brian} and provide a measurement of the spin structure of the nucleon using neutral current elastic scattering.  
I will focus on the CCQE measurement, for other physics topics, please refer to Ref.~\cite{Rex}.

\subsection{Charged-Current Quasielastic Scattering}
The neutrino-nucleus Charged-Current Quasielastic Scattering process ($\nu+n\rightarrow \mu/e+p$ and $\bar{\nu}+p\rightarrow \mu/e+n$)
in the few GeV energy region is very important for neutrino oscillation experiments as it is one of the largest interaction processes and also 
the cleanest detection reaction for both appearance and disappearance searches. The cross section can be described by three dominant form 
factors, two vector form factors $F_{1,2}(Q^2)$ and one axial-vector form factor $F_A(Q^2)$. The vector form factors can be measured from 
electron scattering experiments, and the axial-vector form factor can be measured at non-zero $Q^2$ in neutrino scattering. Assuming a dipole 
form, the $Q^2$ (4-momentum transfer) dependence of the axial-vector form factor can be characterized by the axial mass $M_A$, therefore 
the axial mass can be extracted by fitting the reconstructed $Q^2$ distribution.  
MiniBooNE Collaboration recently reported the first measurement of an absolute $\nu_\mu$ CCQE double differential cross section in muon kinetic 
energy and opening angle using  the world's largest sample of $\nu_\mu$ CCQE events with low-background in the 1 GeV energy region~\cite{MiniBooNE-CCQE}. 
With minimal model dependence, such as Monte Carlo modeling of nuclear effects, the double differential cross section is significantly larger (30\% 
at the flux average energy) than prediction based on relativistic Fermi Gas model (RFG) and the world-average value for the axial mass, 
$M_A=1.03$ GeV~\cite{WA}. A ``shape-only'' fit of the $Q^2_{\rm QE}$ distribution yields an effective axial mass of $M_A=1.35\pm0.17$ GeV, 30\% 
higher than the world average $1.026\pm0.021$ GeV. Using higher energy (3-100 GeV), but same target (Carbon), NOMAD Collaboration reported
the axial mass in a value of $M_A=1.05\pm0.07$ GeV~\cite{NOMAD} which is consistent with the world average. One possible explanation is the 
multi-nucleon emission in the nucleus which will produce a sizable increase in the CCQE cross section~\cite{Martini}. In this scenario, NOMAD may 
only measured a subset ($\mu+p$) that MiniBooNE measured ($\mu+p$ and $\mu+2p$) .
 
\subsubsection{Charged-Current Quasielastic Scattering total cross section}
\begin{figure*}[t]
\centering
\includegraphics[width=135mm]{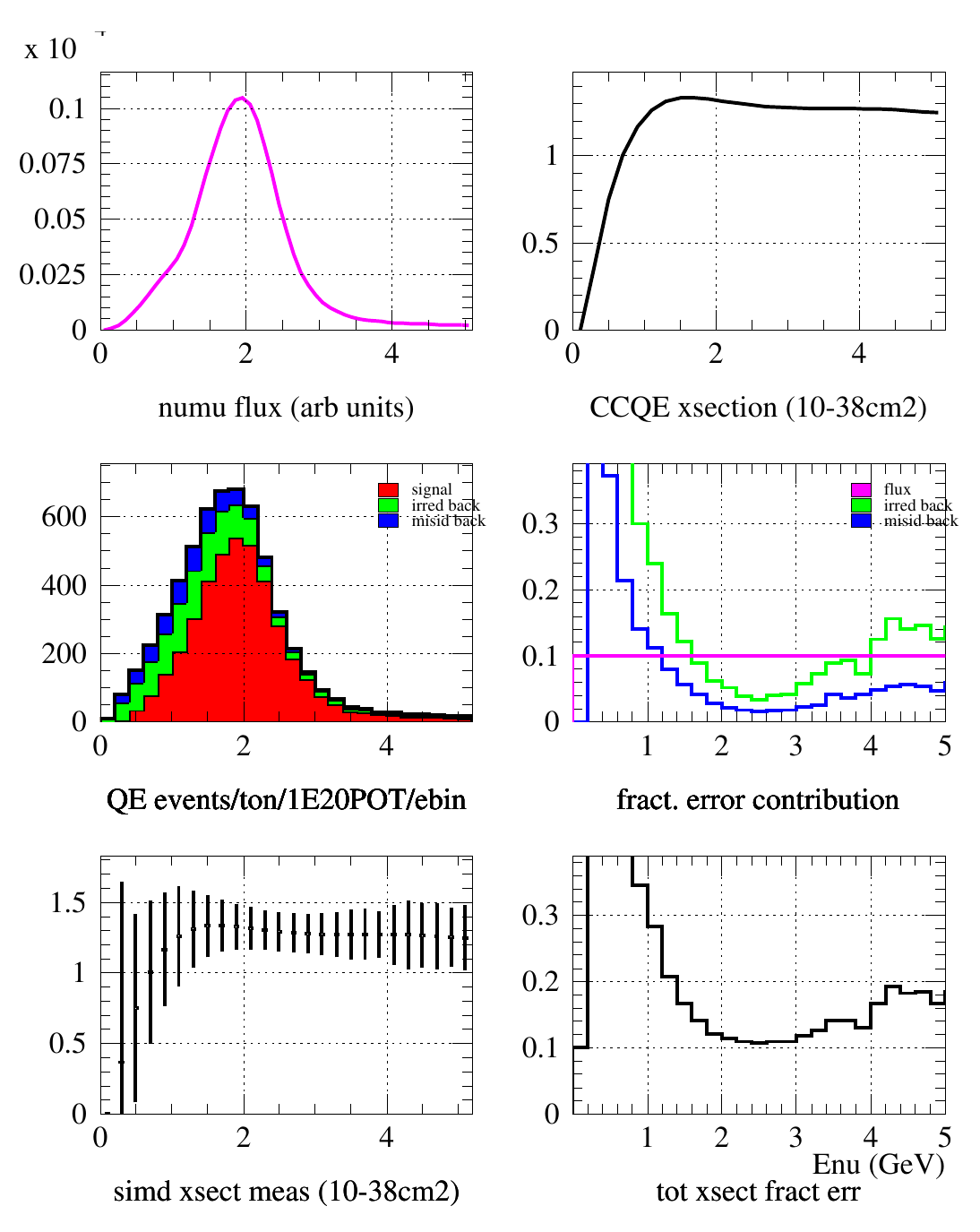}
\caption{Summary of estimated systematic errors on a SciNO$\nu$A total cross section measurement based on 
NUANCE study. The top row shows the neutrino flux and estimated cross section. The middle left plot shows the 
estimated number of accepted CCQE events including signal and two CC background components. The middle 
right plot shows the fractional systematic error on the cross section due to flux and the two CC$\pi^+$ contributions. 
The bottom left plot shows the estimated cross section with the estimated systematic error bars and the bottom right 
shows the corresponding cross section fractional errors. The abscissa on all plots is $E_\nu$, for the bottom two
rows is the reconstructed $E_\nu$ with the assumption of quasielastic kinematics.} \label{ccqe-xs}
\end{figure*}

\begin{figure*}[t]
\centering
\includegraphics[width=135mm]{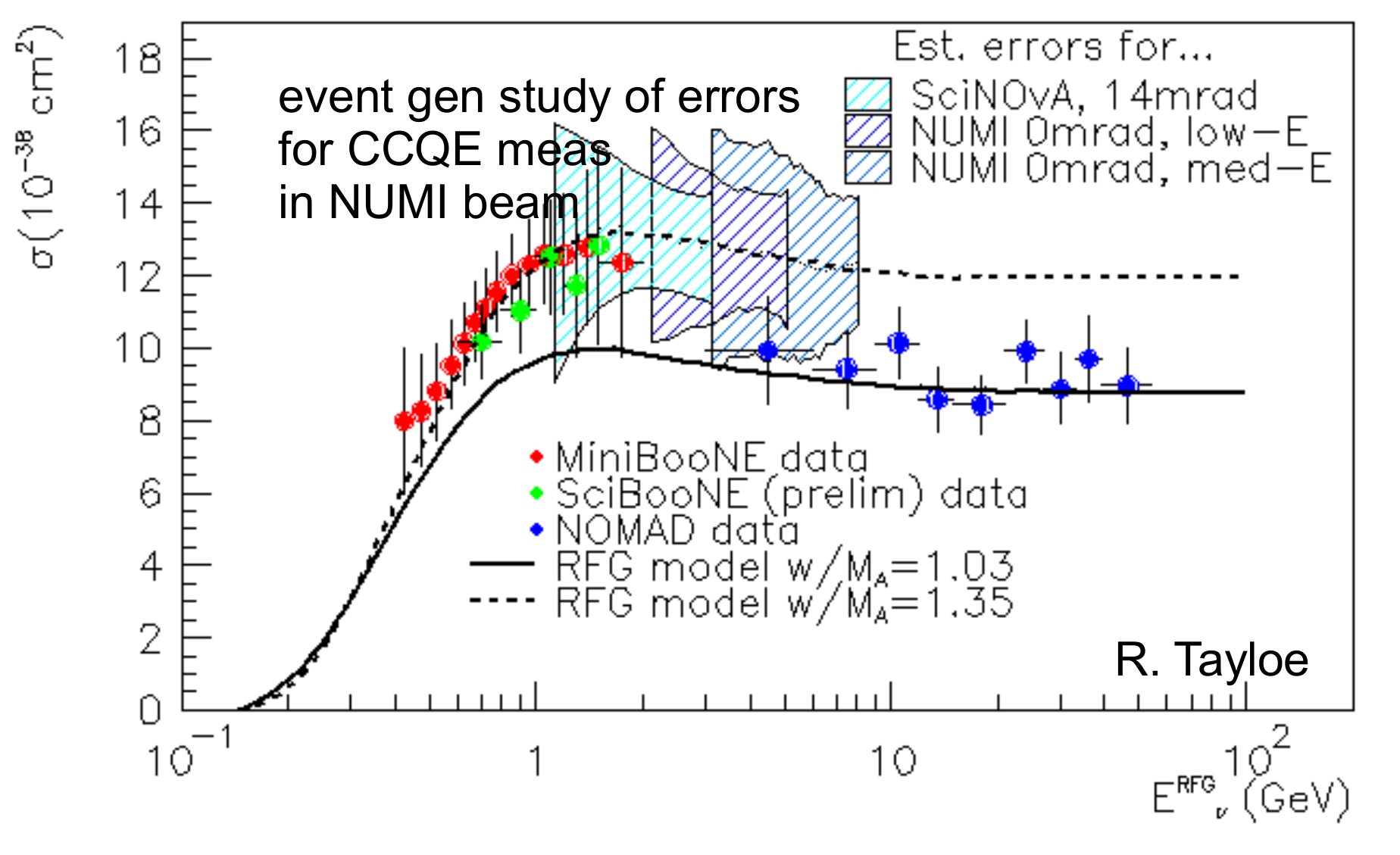}
\caption{A comparison of total errors in $\nu_\mu$ CCQE cross section measurements based on different beam configurations for SciNO$\nu$A 
with recent measurements from MiniBooNE (red dots), NOMAD (blue dots) and SciBooNE (green dots)~\cite{SciBooNE}. The prediction is based 
on relativistic Fermi Gas model with $M_A=1.03$ GeV (solid line) and $1.35$ GeV (dashed line).} 
\label{ccqe-err-com}
\end{figure*}

The dominant systematic uncertainties in the MiniBooNE CCQE measurement are the neutrino flux and charged current
single pion (CC$\pi^+$) production backgrounds. As SciNO$\nu$A will use a narrow band beam, we will have a better prior 
knowledge about the neutrino flux, and also there will be lower CC$\pi^+$ background contamination from the high energy 
feeddown. Following the error analysis of the recent MiniBooNE CCQE analysis, we estimated the systematic errors that 
would result from SciNO$\nu$A measurements of the total cross section. In this analysis, the absolute flux error is assigned 
to be 10\%. The CC$\pi^+$ background contains two categories: pion absorption as modeled by the NUANCE event 
generator~\cite{NUANCE} and pion mis-identification. The results are summarized in Fig.~\ref{ccqe-xs}. The study shows 
that the dominant error in the region of the peak neutrino energy around 2 GeV is the neutrino flux, and the CC$\pi^+$ 
background errors dominate at energies below from the flux peak because of the higher energy ``feed down''. This 10\% 
measurement sitting between the MiniBooNE and NOMAD experiments will add valuable information to the current 
CCQE ``puzzle'' considering the 30\% discrepancy of MiniBooNE and world average. Utilizing the same procedure but 
with the wide band on-axis beam, we found that the errors are $\sim23\%$ for Low Energy (LE) and $\sim35\%$ for 
Medium energy (ME) around 2 GeV which are dominated by the CC$\pi^+$ errors as summarized in Fig.~\ref{ccqe-err-com} 
and Tab.~\ref{ccqe-error}.

\begin{table}[!hbt]
    \begin{center}
        \caption{A summary table of different beam configurations and the corresponding CCQE total cross section errors around 2 GeV.}
        \begin{tabular}{|l|c|}
        \hline        
        {NuMI flux config} & {Estimated error @ 2 GeV (\%)} \\ \hline
        {14 mrad off-axis (SciNO$\nu$A)} & {12} \\ \hline
        {On-axis, LE} & {23} \\ \hline
        {On-axis, ME} & {35} \\ \hline
        \end{tabular}
        \label{ccqe-error}
    \end{center}
\end{table}

\subsubsection{Search for multi-nucleon emission in CCQE interactions}
\begin{figure}[h]
\centering
\includegraphics[width=0.60\textwidth]{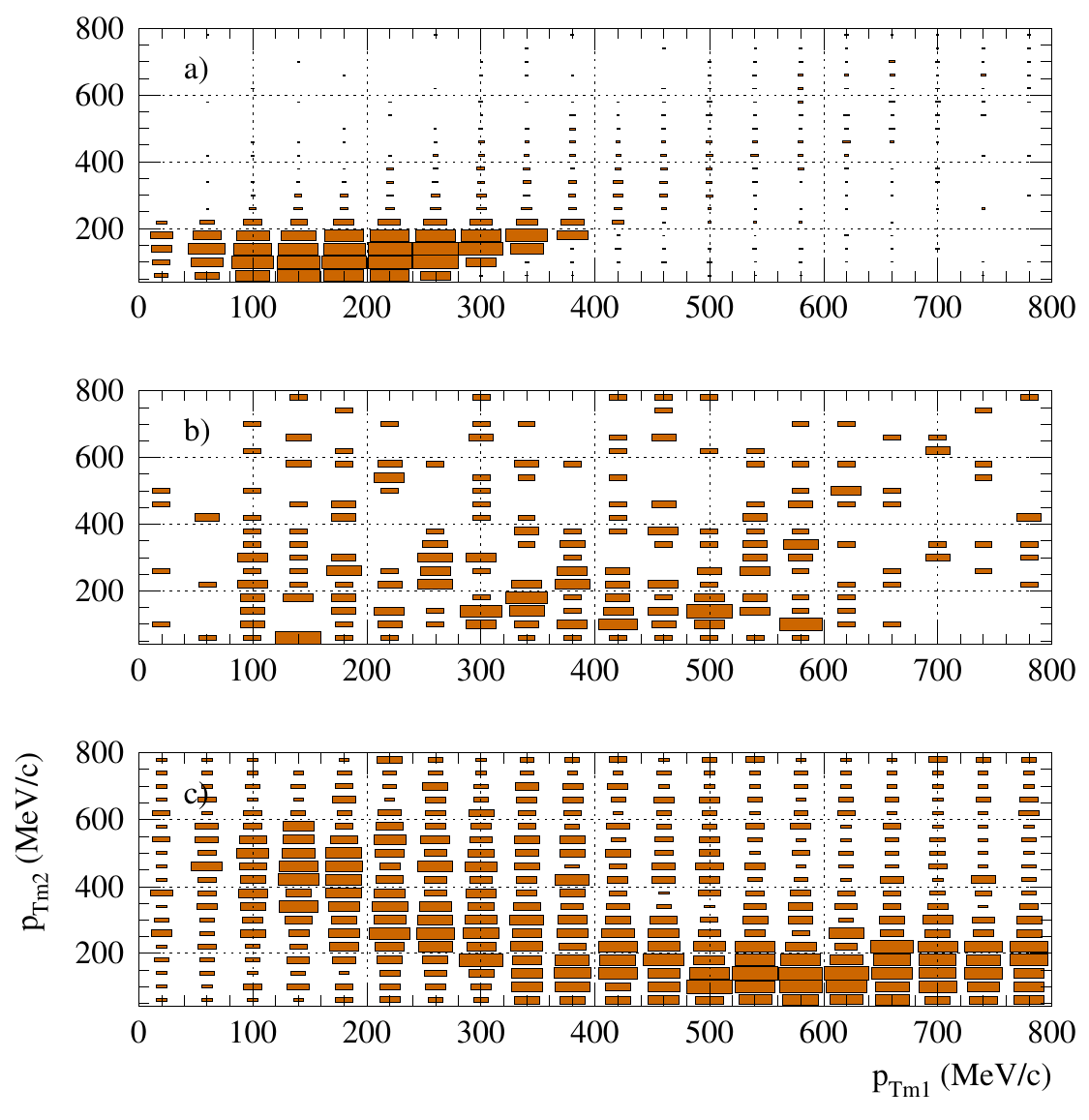}
\caption{The 2D distribution of $P^T_{m1}$ in $y$ versus $P^T_{m2}$ in $x$ passing the selection criteria described in the text. 
The top one is the 2 protons signal; the middle one is CCQE background; and the bottom one is the CC$\pi$ background.} \label{qe-pt}
\end{figure}

\begin{figure}[h]
\centering
\includegraphics[width=0.60\textwidth]{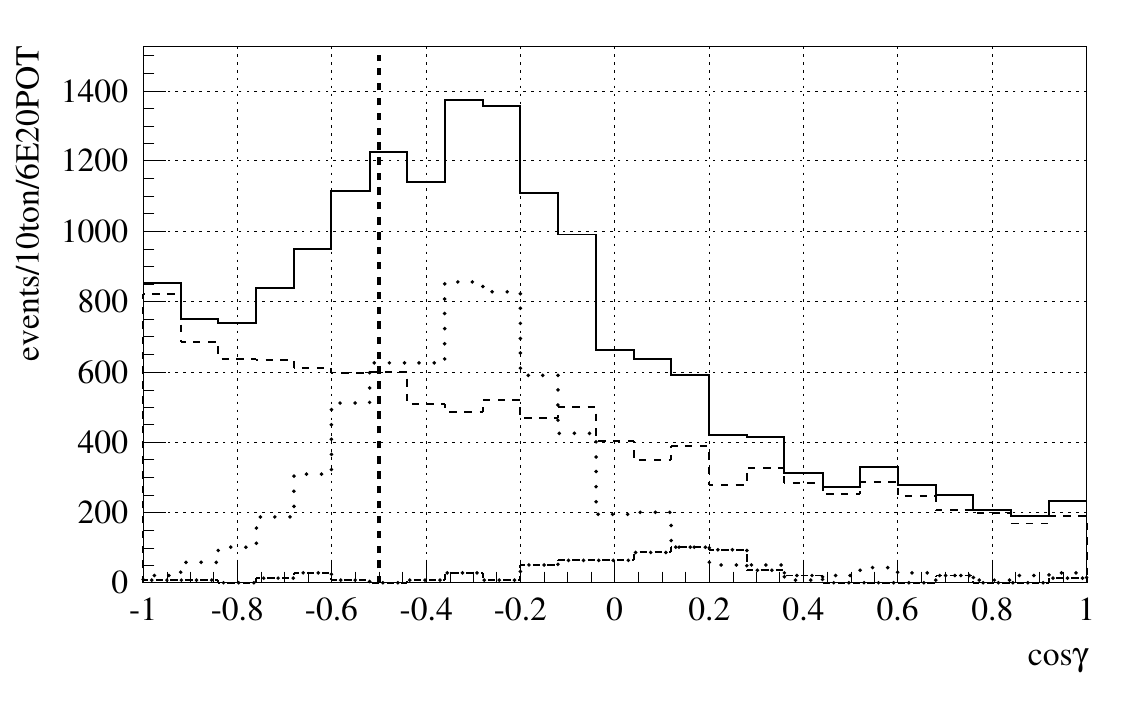}
\caption{The distribution of the opening angle, $\cos\gamma$, between 2 protons passing the selection criteria described in the text.
The 2-nucleon signal is in dashed line, ``true'' CCQE background is in dot-dashed line, CC$\pi$ background is in dotted line, and the 
sum is in solid line. A cut $\cos\gamma<-0.5$ indicated as the vertical dashed line is applied. } \label{qe-angle}
\end{figure}

In the multi-nucleon emission scenario, there is a correlated proton pair in the final state of a CCQE interaction. The fine-grained detector enables us to 
explore the full reconstructed kinematics of two protons, therefore a search for two correlated protons with a muon in the final state can be performed 
in the SciNO$\nu$A detector. The signal is a muon plus two protons, and the backgrounds are from ``true'' CCQE events producing an extra nucleon via final
final state interactions in carbon and CC$\pi^+$ events where the pion is absorbed and multiple nucleons are produced in the final states.  A study has been done using NUANCE 
MC by choosing events with $0.3<Q^2<0.8$ GeV$^2$, corresponding to the bulk of the MiniBooNE measured cross section range. The reconstructed protons 
are required to have a minimum momentum of 450 GeV$/c$. The missing transverse momentum $P^T_{m1}$ calculated from the muon and leading energetic 
proton is required to be greater than 220 MeV/$c$ (Fermi momentum) and $P^T_{m2}$ calculated from the muon and both protons is required to be less than 
220 MeV/$c$. Fig. ~\ref{qe-pt} shows the missing transverse momentum distribution of $P^T_{m1}$ versus  $P^T_{m2}$ for different processes. We also 
require the opening angle between two-proton momentum vectors, $\cos\gamma$, to be less than -0.5 as shown in Fig.~\ref{qe-angle}. After applying all the 
cuts, there are about 4,000 signal events with a signal/background ratio of 3 as shown in Tab.~\ref{ccqe-2n-error}. The fine-grained SciNO$\nu$A detector 
permits us to explore the full reconstructed kinematics which offers rich physics content and should provide an understanding of the underlying 
physics even if 2-nucleon correlations do not prove to be substantial in the CCQE neutrino process. Therefore, SciNO$\nu$A can make a 
significant contribution to the role of 2-nucleons correlations in the CCQE process.

\begin{table}[!hbt]
    \begin{center}
        \caption{The events passing the selection criteria mention above assuming a 10 ton fiducial volume and 6$\times10^{20}$ Protons On Target (POT) per year.}
        \begin{tabular}{|l|c|}
        \hline        
        {Event Type}                                  & {events/10ton/6$\times10^{20}$} \\ \hline
        {2-nucleon signal}                        & {4,119} \\ \hline
        {CCQE 1-nucleon background} & {65} \\ \hline
        {CC$\pi$}                                        & {1,320} \\ \hline
        {Total background}                       & {1,384} \\\hline
        \end{tabular}
        \label{ccqe-2n-error}
    \end{center}
\end{table}

\subsection{Benefit to the NO$\nu$A oscillation measurement}
The primary goal of NO$\nu$A experiment is to search for $\nu_\mu \rightarrow \nu_e$ oscillations using the NuMI neutrino beam. 
The primary backgrounds to this $\nu_e$ appearance search come from NC $\pi^0$ events that can be misidentified as $\nu_e$ signal 
if one of the two photons from the $\pi^0$ decay is not detected and the intrinsic electron-neutrino component of the neutrino beam 
which result from muon and kaon decays. The electron-neutrino charged-current events selection efficiency is $\sim35\%$, while the
acceptance probabilities of neutral-current backgrounds and muon neutrino charged-current events are 0.4\% and 0.1\% respectively.

The background rejection factors for the experiment must not only be large, but they must also be well known to avoid having the sensitivity
of the experiment degraded by uncertainties on the backgrounds. The fine-grained SciNO$\nu$A detector will not only enhance the signal 
efficiency and background rejection estimation, but also add a powerful cross-check of the NO$\nu$A detector performance by 
providing an more efficient signal/background determination which can be compared to the determination made by the NO$\nu$A 
detector, similar in concept to techniques (``Double Scan'') used in the analysis of bubble chamber data. The study shows that  one 
would expect this ``double scan'' technique to find any miscalculations of the signal and background efficiencies as large as a few 
percent. This method is a powerful cross-check of the detector capabilities that uses the neutrino data itself and has very little reliance 
on Monte Carlo simulation and can protect the experiment from unexpected misestimates of background rates.

\begin{figure}[h]
\centering
\includegraphics[width=0.90\textwidth]{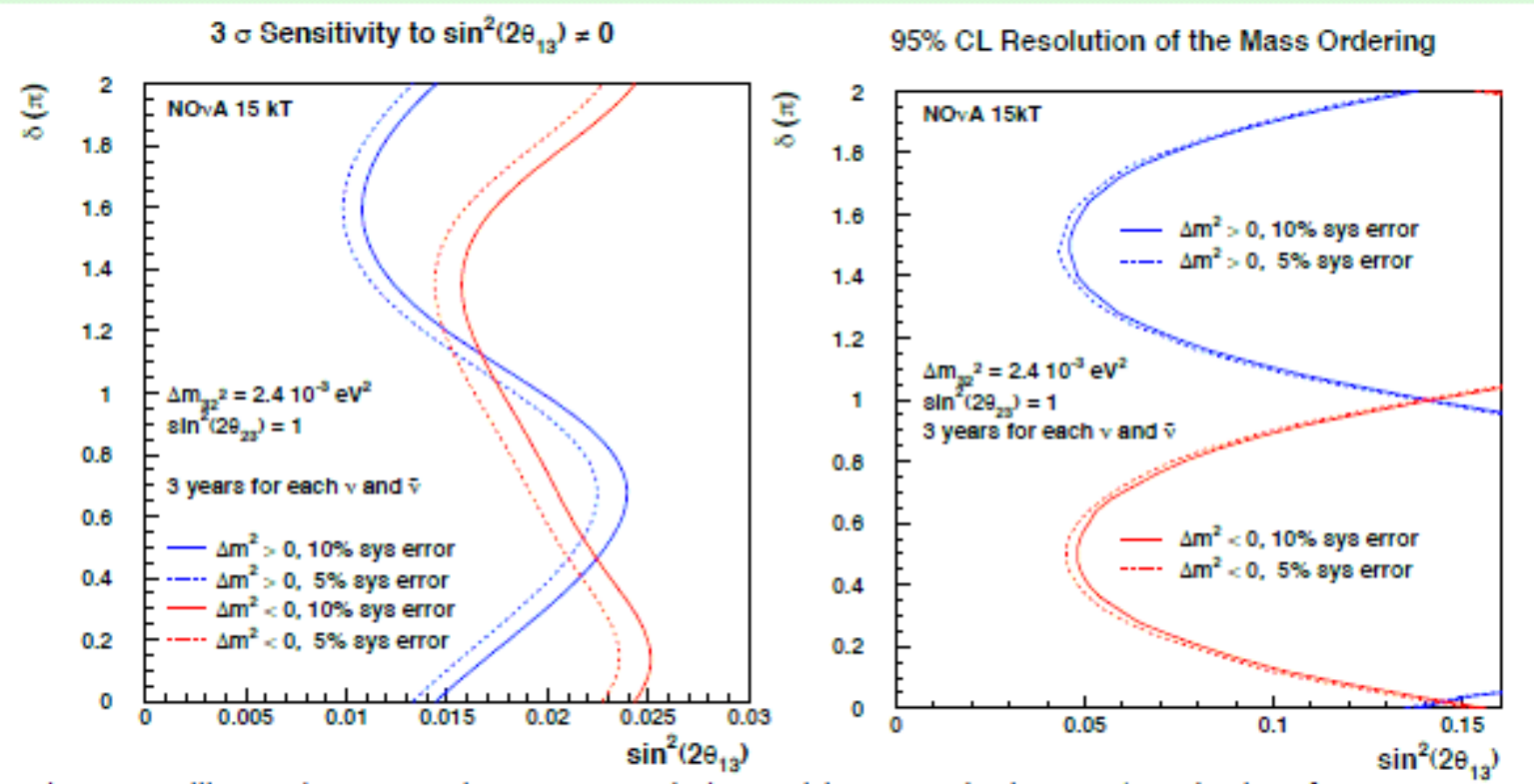}
\caption{The above plots show the SciNO$\nu$A impact on the NO$\nu$A $\sin^2(2\theta_{13})$ sensitivities.
The solid lines assume 10\% uncertainty and the dashed assume 5\% uncertainty in backgrounds. Picking 
one point off the plot, the 3$\sigma$ sensitivity for normal hierarchy and $\delta_{CP}=0$ moves from
0.0144 to 0.0133 - an 8\% improvement. } \label{theta13-imp}
\end{figure}

NO$\nu$A expects a 10\% uncertainty in the background at the far detector. With added data from SciNO$\nu$A it may be possible to 
reduce this uncertainty to 5\% and improve the NO$\nu$A $\sin^2(2\theta_{13})$ sensitivity by 8\% as shown in Fig.~\ref{theta13-imp}. 
To achieve the same reduction in the background uncertainty through increasing the far detector mass would require about 2 additional 
kilotons detector mass which would cost roughly \$20 M.

\begin{figure}[h]
\centering
\includegraphics[width=0.60\textwidth]{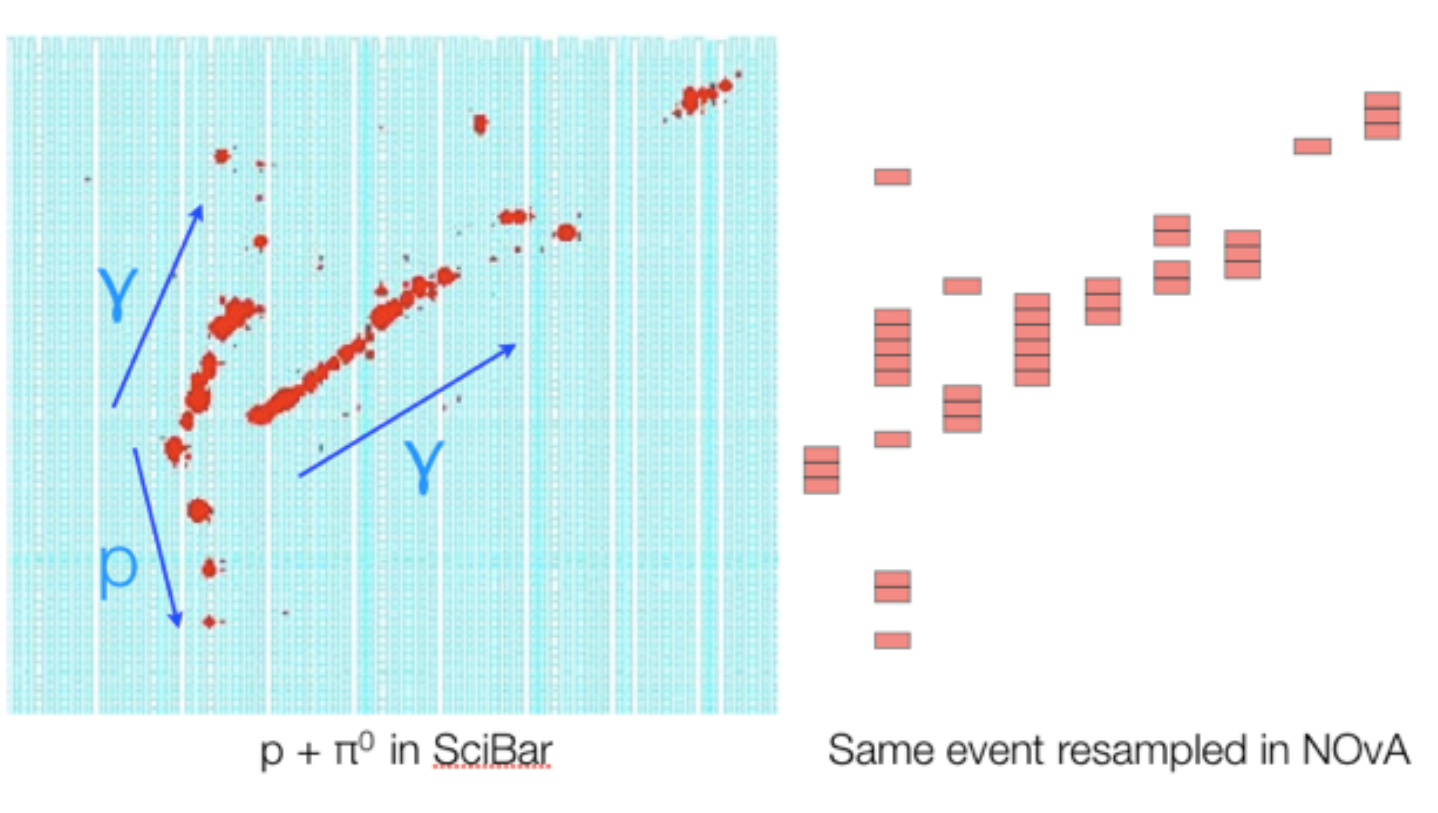}
\caption{Demonstration of the resampling technique. Left panel shows an actual event recorded
by the SciBar detector in the Booster Neutrino Beamline as part of the SciBooNE experiment;
presumably $\nu_\mu+p\rightarrow \nu_\mu+p+\pi^0$. The right panel shows how this same 
event would appear in the NO$\nu$A detector.} \label{double-scan}
\end{figure}

\section{Conclusion}
In summary, SciNO$\nu$A is a proposed experiment to deploy a fine-grained scintillator 
detector in front of the NO$\nu$A near detector using the NuMI, off-axis, narrow-band
neutrino beam at Fermilab. This detector can make unique contributions to the measurement 
of charged- and neutral-current quasi-elastic scattering; charged- and neutral current 
coherent pion production; and neutral-current $\pi^0$ and photon production. These
processes are important to understand for fundamental physics and as backgrounds
to measurements of electron neutrino appearance oscillations. 
\begin{acknowledgments}
The author would like to thank the SciNO$\nu$A study group for the generous help during the presentation and proceeding preparations.
\end{acknowledgments}

\bigskip 

\end{document}